\documentclass[11pt]{article}
\usepackage{amsmath,amssymb,color,graphics,epsfig,cite,CJK}
\usepackage{url}
\usepackage{ulem}
\usepackage[colorlinks=true,citecolor=blue,linkcolor=black]{hyperref}
\usepackage{amsfonts}
\usepackage{subfigure}
\newcommand{\hoch}[1]{$\, ^{#1}$}

\newcommand{\be}{\begin{equation}}
	\newcommand{\ee}{\end{equation}}
\newcommand{\bea}{\setlength\arraycolsep{2pt} \begin{eqnarray}}
	\newcommand{\eea}{\end{eqnarray}}

\def\fft#1#2{{\frac{#1}{#2}}}

\def\0{{\sst{(0)}}}
\def\1{{\sst{(1)}}}
\def\2{{\sst{(2)}}}
\def\3{{\sst{(3)}}}
\def\4{{\sst{(4)}}}
\def\5{{\sst{(5)}}}
\def\6{{\sst{(6)}}}
\def\7{{\sst{(7)}}}
\def\8{{\sst{(8)}}}
\def\sst#1{{\scriptscriptstyle #1}}

\def\del{{\partial}}

\def\M2{ {\bar{\mathcal{M}}}_{(2)} }
\def\Mco2{ {\hat{\mathcal{M}}}_{(D-2)} }

\def\gbar{ {\bar{g} }}
\def\Rbar{{ \bar{R} }}

\def\barnab{{\bar{\nabla}}}

\def\ghat{ {\hat{g} }}
\def\Rhat{{ \hat{R} }}

\def\cK{{\cal{K}}}

\begin{document}
	\begin{CJK}{UTF8}{song}
		\begin{center}
				{\Large {\bf %Revisit
					On Coordinate Singularities Induced by Trapping Horizons
                    
			}}
			
			\vspace{40pt}
			{ Jinbo Yang\hoch{1}}, Hongwei Tan\hoch{2}, Hyat Huang\hoch{3}, and Wen-Cong Gan\hoch{3}
			
			\vspace{8pt}
			
			\hoch{1}{\it Department of Astronomy, School of Physics and Materials Science,\\
				Guangzhou University, Guangzhou 510006, P.R.China}\\
			
			\hoch{2}{\it School of Science, Hunan Institute of Technology, Hengyang 421002, China}\\

            \hoch{3}{\it College of Physics and Communication Electronics, Jiangxi Normal University, Nanchang, 330022, China}
			
			\vspace{40pt}
			
			\underline{ABSTRACT}
		\end{center}
	The trapping (or apparent) horizon serves as a key tool for tracing the complete evolution of black holes. We investigate a class of coordinate singularities induced by such trapping (or apparent) horizons in a spherically symmetric, dynamic spacetime, which are distinct from the well-known coordinate singularities associated with the Killing horizon. In particular, we clarify the geometric structure of this coordinate singularity by means of the Kodama vector field, thereby avoiding unphysical artifacts. We further employ the evolving Ellis drainhole as an analytical model to illustrate key details of this phenomenon.
		
	\vfill {\footnotesize %yangjinbo@gzhu.edu.cn, \par
Corresponding author: honweitan@hnit.edu.cn     
    }
		\thispagestyle{empty}
		
		\pagebreak
		\tableofcontents
		\addtocontents{toc}{\protect\setcounter{tocdepth}{2}}
	
		%%%%%%%%%%%%%%%%%%%%%%%%%%%%%%%%%%%%%%%%
		
		\newpage	
		
		%%%%%%%%%%%%%%%%%%%%%%%%%%%%%%%%%%%%%%%%

%%%%%%%%%%%%%%%%%%%%%%%%%%%%%%%%%%%%%
\section{Introduction}
\setcounter{equation}{0}
\renewcommand\theequation{1.\arabic{equation}}
Understanding the evolution of black holes has emerged as a critical challenge. From an observational perspective, the abundant ``little red dots" recently discovered by the James Webb Space Telescope appear to demand a deeper comprehension of the seeding and growth of supermassive black holes \cite{Matthee:2023utn}.
From a theoretical physics standpoint, Hawking evaporation \cite{Hawking:1976ra} gives rise to the renowned information paradox. Black holes emit thermal radiation that diminishes their mass, thereby reducing their event horizon area, i.e., a violation of the Null Energy Condition (NEC). Once a black hole evaporates entirely, only thermal radiation remains, carrying no information about the state of the matter that collapsed to form it. This process is fully described from the perspective of an observer outside the black hole horizon, who witnesses the horizon first form and subsequently vanish.
Thus, to describe this process accurately, we must clarify key notions regarding black hole horizons and adopt appropriate tools to track horizon evolution. We further show that NEC violation is a direct consequence of the horizon being observable to external observers.

The so-called trapping horizon or apparent horizon serves as such a tool. First proposed by Hayward \cite{Hayward:1997jp}, a trapping horizon is defined as a hypersurface foliated by marginally trapped surfaces (MTSs) or marginally anti-trapped surfaces (MATSs). Notably, MTSs and MATSs share a unifying characteristic: they correspond to orientable closed surfaces penetrated orthogonally by a congruence of null geodesics with vanishing expansion.
As a general concept encompassing multiple subtypes, the trapping horizon finds applications beyond black holes \cite{Hayward:1993wb, Hayward:1997jp, Hayward:1998pp, Hayward:2005gi, Harada:1998wb, Harada:2013epa, Hioki:2019gnv, Saida:2007ru, Sato:2022yto, Binetruy:2014ela, Helou:2016xyu, Binetruy:2018jfz}, extending to white holes \cite{Harada:2021xze}, Hubble horizons in cosmological contexts \cite{Cai:2005ra, Cai:2006rs, Akbar:2006er, Akbar:2006kj, Sheykhi:2007zp, Cai:2008mh, Cai:2008ys, Cai:2008gw}, and even traversable wormholes \cite{Hochberg:1998ii, Shinkai:2002gv, Hayward:2009yw, Maeda:2009tk, Tomikawa:2015swa, Bittencourt:2017yxq, Yang:2021diz, Liang:2025lay}. For black hole scenarios specifically, MTSs rather than MATSs should be employed, with the congruence of null geodesics propagating outward. For generality, we do not specify the type of trapping horizon in this work.

It is also noteworthy that the terminology ``apparent horizon" (as practically utilized) coincides with the trapping horizon in black hole contexts, yet deviates from its original definition \cite{HawkingEllis, WaldGR}. Specifically, in four-dimensional spacetime, the trapping horizon is a three-dimensional hypersurface, whereas the originally defined apparent horizon is a two-dimensional surface dependent on the choice of Cauchy slicing \cite{Wald:1991zz, Faraoni:2016xgy}. For the precise definition of the original apparent horizon, we refer readers to \cite{HawkingEllis, WaldGR}. Nevertheless, these concepts share a core feature: vanishing expansion. This property enables a quasi-local characterization of the black hole boundary, overcoming the teleological limitation inherent in treating the event horizon as the black hole boundary \cite{HawkingEllis, WaldGR}. Throughout this paper, ``horizon" refers exclusively to the trapping horizon or the practically employed apparent horizon, not the original apparent horizon or the event horizon.

Along similar lines, a series of studies has advanced our understanding of the implications of horizon observability and explored universal near-horizon properties \cite{Baccetti:2018otf, Terno:2019kwm, Terno:2020tsq, Murk:2020ftq, Dahal:2021hbm, Mann:2021lif, Murk:2021qiv, Murk:2021qdb, Dahal:2022pig, Murk:2022dkt, Terno:2022qot, Murk:2023rwl, Murk:2023vdw, Dahal:2023hzo}.
Notable conclusions have emerged from this body of work, including: the exclusion of the single collapsing dust model for the formation of observable black holes \cite{Terno:2019kwm}; the emergence of a firewall (as a weak singularity) at the outer boundary of the trapped region \cite{Terno:2019kwm, Terno:2020tsq, Dahal:2021hbm}; the inaccessibility of traversable wormholes \cite{Terno:2022qot}; and subtleties in generalizing surface gravity definitions for dynamical black holes \cite{Mann:2021lif}. This approach has further been extended to modified gravity \cite{Murk:2020ftq, Murk:2021qiv, Murk:2021qdb, Dahal:2022pig, Murk:2022dkt}, regular black holes \cite{Murk:2023rwl, Murk:2023vdw}, and black holes embedded in cosmological backgrounds \cite{Dahal:2023hzo}, offering valuable insights into black hole evolution.
This series of studies builds on the core results initially presented in \cite{Baccetti:2018otf}.

We summarize the key results of \cite{Baccetti:2018otf} herein. Their approach explores universal near-horizon features in dynamically evolving, spherically symmetric spacetimes, without being restricted to spherical black holes. They adopt the $\{t,r,\theta,\phi\}$ coordinate system, which allows the spacetime line element to be written as
\be 
ds^2 = - e^{2h(t,r)} f(t,r) dt^2 + \frac{dr^2}{f(t,r)} + r^2 d\Omega^2 \label{orthSph} \,, 
\ee
where $h(t,r)$ and $f(t,r)$ are functions of $t$ and $r$. The absence of a $dtdr$ term implies that time slices labeled by $t$ are orthogonal to constant-$r$ surfaces. The dynamical nature of the spacetime is characterized by the $t$-dependence of $h$ and $f$. Moreover, the horizon location $r=r_g(t)$ is determined by the equation $f(t,r_g(t))=0$.
\begin{align} 
  \lim_{r\rightarrow r_g} e^{-2h}T_{tt} 
= \lim_{r\rightarrow r_g} T^{rr} 
= \Xi \,, \quad \lim_{r\rightarrow r_g} e^{-h}T^{r}_{\;\,t} 
= \pm \Xi \,. \label{EMTintrCommonLimit} 
\end{align}
This reveals a universal near-horizon structure in the energy-momentum tensor (EMT):

\begin{align} T_{ab} = \left(\begin{array}{ccc} e^{2h}\,\Xi & \pm e^{h}\,\Xi/f \\ \pm e^{h}\,\Xi/f & \Xi f^{-2} \end{array}\right) \,, \label{EMTintrnearhorizon} \end{align}

where $a, b$ are indices for the pre-two-dimensional sub-spacetime, corresponding to the $t$ and $r$ directions. Furthermore, Ref. \cite{Baccetti:2018otf} introduces the function $C=r(1-f)$, which equals twice the Misner-Sharp (MS) mass in Einstein gravity with geometric units \cite{Hayward:1994bu}.

They then introduce the following coordinate transformation to analyze near-horizon behavior: $t_x=t$, $x = r - r_g(t)$, and $W=C-r_g(t)$. Under a suitable regularity condition for the horizon, they derive

\begin{align} \frac{\partial W}{\partial x} \approx \frac{8\pi \Xi \,r_g^3}{x-W} \,, \label{MMSpriintx} \end{align}

where $\approx$ denotes the dominant term as $f\rightarrow 0$. Regularity also imposes constraints on the function $h$:

\begin{align} \frac{\partial h}{\partial x} \approx \frac{8\pi \Xi \,r_g^3}{(x-W)^2} \,. \label{hpriintx} \end{align}

The approximate near-horizon solutions to Eqs. \eqref{MMSpriintx} and \eqref{hpriintx} characterize the near-horizon geometry, forming the core results of Ref. \cite{Baccetti:2018otf}.

However, the horizon location $r=r_g(t)$ introduces an explicit coordinate singularity in the line element \eqref{orthSph} as $f\rightarrow 0$. Does this coordinate singularity imply unphysical consequences? We demonstrate that constant-$t$ surfaces are tangent to constant-$r$ surfaces at the horizon. This geometric configuration gives rise to the coordinate singularity, as the cotangent vector fields $dt$ and $dr$ become collinear and thus fail to form a valid coordinate basis. To address this singularity, we propose a covariant approach based on the Kodama vector field \cite{Kodama:1979vn}. Additionally, we introduce the Ellis drain hole as a tractable example \cite{Ellis:1979bh} to verify and illustrate key details of this covariant method.

This paper is organized as follows. In Section 2, we clarify the geometric origin of the coordinate singularity induced by $f=0$ using the Kodama vector field. This clarification also provides a covariant framework to reproduce Eqs. \eqref{EMTintrCommonLimit}, \eqref{MMSpriintx}, and \eqref{hpriintx} under a more relaxed horizon regularity condition. Section 3 introduces the Ellis drain hole and analytically constructs a coordinate transformation to the $\{t,r,\theta,\phi\}$ system. The analytic form of this transformation enables rigorous analysis of the functions $h(t,r)$ and $f(t,r)$, as well as comparisons of EMT components across different coordinate systems. Section 4 summarizes the main conclusions and discusses their implications for understanding black hole evolution.

%%%%%%%%%%%%%%%%%%%%%%%%%%%%
\section{General consideration}
\setcounter{equation}{0}
\renewcommand\theequation{2.\arabic{equation}}
This section clarifies the geometric picture underlying the coordinate singularity in Eq.\eqref{orthSph} based on the Kodama vector field introduced in Ref. \cite{Kodama:1979vn}.
Firstly, we do not specify a particular coordinate system for a four-dimensional spherically symmetric spacetime, but instead consider the following general line element:
\begin{align}
	ds^2 = \gbar_{ab}\, du^a du^b + r^2(u)\, d\Omega^2 \,,
\end{align}
where $d\Omega^2$ denotes the unit 2-sphere, $u^a$ are arbitrary coordinates $\{u^0,u^1\}$ for the two-dimensional sub-spacetime $\M2$, which possesses an independent metric with components $\gbar_{ab}$. The scalar function $r(u)$ on $\M2$ is the areal radius. We also denote the covariant derivative for $\M2$ as $\barnab_{a}$, which satisfies $\barnab_a \gbar_{bc}=0$. For brevity, we define $\barnab^{a}r = \gbar^{ab} \barnab_b r$.

Next, we assume evolving horizons are regular. Specifically, this assumption requires the metric of $\M2$ and the areal radius $r(u)$ to be smooth in the neighborhood of horizons. Thus, regularity demands that $r(u)$, $\barnab^{a}r$, and $\barnab_{a}\barnab_{b}r$ are continuous tensors near horizons. Employing a horizon-covered coordinate system, the components of these tensors must be finite at the horizon.
We emphasize that the full regularity requirement is necessary. Suppose components of the energy-momentum tensor (EMT) projected onto a normalized orthogonal tetrad diverge, while several scalar combinations of the EMT remain finite. Such scenarios also correspond to geodesic incompleteness \cite{HawkingEllis, WaldGR, Elizalde:2004mq}, analogous to examples of big-rip singularities in certain dark energy models \cite{Elizalde:2004mq}.

Although we only assume spherical symmetry and regular evolving horizons, introducing two future-directed smooth null vector fields $k^{a}$ and $l^a$ facilitates capturing key features of the near-horizon geometry. We require $k^{a}$ and $l^a$ to satisfy the normalization condition $\gbar_{ab} k^{a} l^{b}=-1$, forming a double null tetrad $\{k^{a}, l^{a}\}$ on $\M2$.
Their expansions are then calculated directly as:
\begin{align}
	\theta_{k} = \frac{2}{r}\, k^a \barnab_a r  \,,\quad
	\theta_{l} = \frac{2}{r}\, l^a \barnab_a r  \,, \label{expansionsTrick}
\end{align}
Together with $k_a l^a=-1$, these formulas yield an expression for $\barnab^a r$ in terms of $k^a$ and $l^a$:
\begin{align}
	\barnab^a r = -\frac{r}{2} \big( \theta_{k}\, l^a + \theta_{l}\, k^a \big)  \,.
\end{align}
Note that $\theta_{k}=0$ and $\theta_{l}=0$ determine the locations of all trapping horizons, where $\barnab^a r$ becomes a null vector. We then define the function $f$ (matching the inverse metric component $g^{rr}$ from the line element \eqref{orthSph}) as:
\begin{align}
	f \equiv% g^{\mu\nu} (\nabla_{\mu} r) (\nabla_{\nu} r) = 
	\gbar^{ab} (\barnab_a r) (\barnab_b r) 
	= - \frac{r^2}{2} \theta_{k}\theta_{l}
	\,,
\end{align}
where we also express $f$ in terms of the expansions $\theta_{k}$ and $\theta_{l}$. The roots of the equation $f=0$ thus indicate the locations of horizons.

%%%%%%%% Kodama vector %%%%%%%%%
The so-called Kodama vector is crucial for our covariant approach. It acts as a tangent vector field on $\M2$, defined via the Hodge dual of $\barnab_a r$ as follows:
\begin{align}
	K^a \equiv  
	-\bar{\epsilon}^{ab}  \barnab_b r  \,,
\end{align} 
where $\bar{\epsilon}^{ab}$ is the volume two-form $\bar{\epsilon}_{ab}$ for $\M2$ after index raising. Within the double null tetrad, we specify $\bar{\epsilon}^{ab}$ as
\begin{align}
	\bar{\epsilon}^{ab} = l^a k^b - k^a l^b \,.
\end{align} 
Hence, the Kodama vector expressed in terms of the double null tetrad is
\begin{align}
	K^a = \frac{r}{2} \big( \theta_{l} k^a - \theta_{k} l^a \big)  \,.\label{KodamaExpbykl}
\end{align}
The vector $K^a$ is tangent to every surface of constant $r$, as $K^a\barnab_a r= \bar{\epsilon}^{ba} \barnab_a r\barnab_b r =0$. The length of $K^a$ is determined by its inner product with itself:
\begin{align}
	\gbar_{ab} K^a K^b = -f  \,.  \label{KodamaIPro}
\end{align} 
Consequently, $K^a$ is also null on the horizon, where $K^a$ and $\barnab^a r$ become collinear. For instance, if the horizon is determined by $\theta_{k}=0$, Eq.\eqref{KodamaExpbykl} gives $[\barnab^a r]_H = -[K^a]_H $, where the subscript $H$ denotes evaluation on the horizon. Another scenario with $\theta_{l}=0$ yields $[\barnab^a r]_H = [K^a]_H $. These features can be summarized as $[K^a]_H=\pm [\barnab^a r]_H$ \footnote{We retain the same subscript $H$ without specifying $\theta_{k}=0$ or $\theta_{l}=0$.}, a property analogous to that of Killing vector fields in static spacetimes.

\begin{figure}[htbp]%[htbp!][H]
	\begin{center}
		\includegraphics[width=4.3cm]{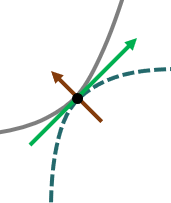}
	\end{center}
	\caption{\small  The gray curves represent surfaces of constant $r$, and the muted teal dashed curve represents a time slice, which is orthogonal to every surface of constant $r$.
	These curves become tangent to each other at the MTS represented by the black dot. The green arrow denotes a future-directed null vector with vanishing expansion. For instance, if $\theta_{k}=0$, the green arrow represents $k^a$; if $\theta_{l}=0$,it represents $l^a$. In contrast, the brown arrow denotes another future-directed null vector.
	} 
\label{Ktangentrconst}  
\end{figure}

%%%%%% The picture %%%%%%
The collinearity of $K^a$ and $\barnab^a r$ on the horizon is a general feature, which signals the coordinate singularity of the $\{t,r\}$ system as $f\rightarrow 0$. Note that this paper focuses on evolving horizons, which do not possess a constant areal radius $r$. Hence, the horizon must be penetrated by a sequence of constant-$r$ surfaces.
On the other hand, the coordinate system $\{t,r\}$ defines time slices labeled by $t$ that are orthogonal to constant-$r$ surfaces. This orthogonality condition $\barnab^a t \barnab_a r=0$ further implies $\barnab^a t \propto K^a$, meaning the Kodama vector can be interpreted as the normal vector to the time slices. Therefore, the collinearity $[K^a]_H=\pm [\barnab^a r]_H$ implies that time slices also penetrate the horizon and are tangent to a constant-$r$ surface there. Regardless of how $t$ is rescaled, the normal vectors $\barnab^a t$ and $\barnab^a r$ fail to form a suitable basis for tangent vectors on the horizon, unless the orthogonality condition is abandoned.
This constitutes the physical picture underlying the coordinate singularity arising from $f\rightarrow 0$ in the $\{t, r\}$ system, as illustrated in Fig.\ref{Ktangentrconst}. The gray curve corresponds to a constant-$r$ surface, and the dashed curve represents a time slice. They are tangent to each other at the black dot, which denotes a cross section of the horizon (i.e., a MTS). The green arrow indicates the null direction with vanishing expansion, while the brown arrow denotes the other null direction. This picture motivates a covariant approach to directly derive Eqs.\eqref{EMTintrCommonLimit}, \eqref{MMSpriintx}, and \eqref{hpriintx}, and justifies the general validity of Eq.\eqref{EMTintrnearhorizon}.
%%%%%%%%%%%%%%%%%%%%%%%%%%%%%%%%%%
It is worth noting that the above scenario differs significantly from that of a Killing horizon or a non-expanding horizon, where the horizon is a null hypersurface. Nullity requires the horizon to correspond to a specific surface with constant $r$, such that no slice can penetrate the horizon without abandoning the orthogonality condition.

%%%%%%%%%%%%%  Near horizon  %%%%%%%%%%%%%%%%%%%%%% 
We then verify the common limit in Eq.\eqref{EMTintrCommonLimit} and the results in Eqs.\eqref{MMSpriintx} and \eqref{hpriintx} based on $[K^a]_H=\pm [\barnab^a r]_H$. We can specify the direction and length of the Kodama vector $K^a$ to calculate its components in the $\{t,r\}$ coordinate system. Recall that $K^a$ is tangent to every surface of constant $r$, so it should only possess a time component $K^{t}$. Meanwhile, its length is determined by the inner product in Eq.\eqref{KodamaIPro}. Hence, the time component is $K^{t} = e^{-h}$ to satisfy this inner product.\footnote{We ignore the sign of $K^{t}$ since it carries no essential distinction. One can always adjust the choice of $t$ to retain $K^{t} = e^{-h}$ rather than $K^{t} = -e^{-h}$.} Transforming to arbitrary coordinates, we obtain
\begin{align}
	K^a = e^{-h} \frac{\partial u^a}{\partial t} \,. \label{Kodamaintrsys}
\end{align}
We then consider a regular symmetric tensor $T_{ab}$. For adjusted components such as $e^{-2h} T_{tt}$ and $e^{-h}T^r_{\;\,t}$ discussed in Refs.\cite{Baccetti:2018otf}, Eq.\eqref{Kodamaintrsys} yields the following relations:
\begin{align}
	&e^{-2h} T_{tt} \equiv T_{ab} K^aK^b  \,, \qquad
	e^{-h}T^r_{\;\,t} \equiv  T_{ab} K^a (\barnab^b r)  \,.
\end{align}
We have rewritten $e^{-2h} T_{tt}$ and $e^{-h}T^r_{\;\,t}$ in covariant form. The terms $T_{ab} K^aK^b$ and $T_{ab} K^a (\barnab^b r)$ must be finite on the horizon, as $T_{ab}$, $K^a$, and $\barnab_a r$ are regular there. Regularity also implies continuity, so the near-horizon limits of $e^{-2h} T_{tt}$ and $e^{-h}T^r_{\;\,t}$ should coincide with $[T_{ab} K^aK^b]_H$ and $[T_{ab} K^a (\barnab^b r)]_H$, respectively.
Recall that $T^{rr} \equiv T^{ab} (\barnab_a r) (\barnab_b r) $ is already a covariant expression. From $[K^a]_H=\pm [\barnab^a r]_H$, we find
\begin{align}
	&[T_{ab} K^aK^b]_H  = 
	\pm[T_{ab} K^a (\barnab^b r)]_H = [T^{rr}  ]_H  \,, \label{comlimCovApp}
\end{align}
which confirms Eq.\eqref{EMTintrCommonLimit} by identifying $\Xi\equiv [T^{rr}]_H$.

Moreover, the physical significance of $\Xi$ can be established by applying the Einstein field equations $G_{\mu\nu}=8\pi T_{\mu\nu}$. Note that $G^{rr}\equiv G_{\mu\nu} \nabla^{\mu}r\nabla^{\nu}r\equiv G_{ab} \barnab^{a}r\barnab^{b}r$. We have
\begin{align}
	G^{rr}=& -\frac{2}{r}\, \big(\barnab^a r\big)\big(\barnab^b r\big) \big(\barnab_a \barnab_b\, r \big) %\nonumber\\&
    + f\,\bigg( \frac{2}{r}\barnab^2r -\frac{1-f}{r^2} \bigg) 
	\,,\label{EinTensor_rr}
\end{align} 
where $\barnab^2r$ denotes the Laplacian of $r$ on $\M2$. On the horizon, the second term of $G^{rr}$ vanishes, as $\barnab^2r$ is finite and $f=0$ there. From $[G^{rr}]_H=8\pi \Xi$, we obtain
\begin{align}  
	\Xi = -\bigg[\frac{(\barnab^ar ) (\barnab^br ) ( \barnab_a\barnab_br  )}{4\pi\,r} \bigg]_H  \,. \label{defXi}
\end{align}
%%%%%%%%%%%%%%%%%%%%%%%%%%%%%%%%%%%%%%%%%%
Without loss of generality, we discuss the horizon defined by $\theta_{k}=0$. Consequently, $[\barnab^a r]_H$ is proportional to $k^a$, such that $\Xi \propto - [k^ak^b( \barnab_a\barnab_br  )]_H $.
The null vector field $k^a$ generates a null geodesic congruence, though its parameter may not be affine. However, $k^a$ can be appropriately rescaled as $ N^a = \beta^{-1} k^a $ to ensure $ N^a$ generates geodesics with an affine parameter, i.e., $N^a \bar{\nabla}_a  N^b=0$. Denoting the affine parameter as $\lambda$, we find $N^a N^b \bar{\nabla}_a \bar{\nabla}_b r = N^a \bar{\nabla}_a \big( N^b\bar{\nabla}_b r \big) = d^2r/d\lambda^2 $. Therefore, 
\begin{align}  
	\Xi	\propto -\bigg[\frac{d^2r}{d\lambda^2}\bigg] _H	\,,
\end{align}
which relates the NEC to the sign of the second derivative of $r$ with respect to $\lambda$. Hence, violation of the NEC corresponds to a minimum of $r$ along the null geodesics generated by $k^a$.

%%%%%%%%%%%%%%%%%%%
We then derive Eq.\eqref{MMSpriintx} and Eq.\eqref{hpriintx}.
It is important to emphasize that the transformation rules for the coordinate basis induced by $t_x=t \,,\quad  x = r - r_g(t) $ must include
\begin{align}
	\frac{\partial}{\partial t} = \frac{\partial}{\partial t_x} - \dot{r}_H  \frac{\partial}{\partial x} \,,\quad  \frac{\partial}{\partial r}  = \frac{\partial}{\partial x} 	\,,
\end{align}
such that $\partial W/\partial x = \partial W/\partial r = \partial C/\partial r $ and $\partial h/\partial x = \partial h/\partial r$. Additionally, $r-C=x-W$ holds since $\partial r_g(t)/\partial x=0$. Using $\partial/\partial r$ rather than $\partial/\partial x$ suggests a covariant approach to deriving Eqs.\eqref{MMSpriintx} and \eqref{hpriintx}.
Recall that $ \gbar_{ab}( \partial u^b/\partial r) = \barnab_a r/f $ and $f=\barnab^a r \barnab_a r $. We rewrite $\partial f/\partial r$ in covariant form as:
\begin{align}
	\frac{\partial f}{\partial r} 
	= \frac{ \barnab^a r \barnab_a f }{f}
	=&~ \frac{2}{f} \,(\barnab^ar ) (\barnab^br ) ( \barnab_a\barnab_br  )	\,.
\end{align}
Consider the function $C=r(1-f)$, which equals twice the Misner-Sharp (MS) mass in four-dimensional general relativity. Its partial derivative with respect to $r$ is:
\begin{align}
	\frac{\partial C}{\partial r} 
	=&~ 1-f -r \frac{\partial f}{\partial r} %\nonumber\\
	\approx  \frac{8\pi \Xi \, r_g^3}{r-C} 
	\,, \label{MMSpriintr}
\end{align}
where we used the relation $1/f\approx r_g/(r-C)$. The regularity of $\barnab_a\barnab_br$ ensures that $\partial C/\partial r$ is dominated by the third term as $f\rightarrow 0$. Thus, Eq.\eqref{MMSpriintr} confirms Eq.\eqref{MMSpriintx}.

On the other hand, $(dK)_{ab}$ must be a regular two-form since $K_a$ is regular. Recall that $K_a=-e^h f \barnab_a t$, we calculate $(dK)_{ab}$ as follows: 
\begin{align}
	(dK)_{ab} 
	=&~ \bigg( f\frac{\partial h}{\partial r} + \frac{\partial f}{\partial r}\bigg) \bar{\epsilon}_{ab} 	\,,
\end{align} 
where $\bar{\epsilon}_{ab} = e^h (dt\wedge dr)_{ab}$ in the $\{t,r\}$ coordinate system.
Regularity implies $\partial h/\partial r$ takes the form:
\begin{align}
	\frac{\partial h}{\partial r}= &~ -\frac{1}{f} \bigg(\frac{\partial f}{\partial r} - \frac{\bar{\epsilon}^{ab}(dK)_{ab}}{2} \bigg)  
	\approx \frac{8\pi \Xi \,r_g^3}{(r-C)^2} 
	\,. \label{parhbyparrEq}
\end{align}
Again, Eq.\eqref{parhbyparrEq} matches Eq.\eqref{hpriintx} due to $\partial h/\partial r=\partial h/\partial x$.
%%%%%%%%%%%%%%%%%%%%%%%%

Nevertheless, such a common limit arises from the coordinate singularity of the $\{t,r,\theta,\phi\}$ system, rather than the intrinsic nature of $T_{ab}$. As an example, we examine $T_{ab}$ on the horizon $\theta_{k}=0$. Define $T_{kk}\equiv T_{ab}k^a k^b$, $T_{ll}\equiv T_{ab}l^a l^b$, and $T_{kl}\equiv T_{ab}k^a l^b$. We then expand $T_{ab}$ as 
\begin{align}
	T_{ab}  =  T_{kk}l_a l_b + T_{kl} (k_al_b+l_ak_b) + T_{ll} k_ak_b \,.
\end{align} 
The limit \eqref{comlimCovApp} thus retains only the $T_{kk}$ component, while neglecting $T_{kl}$ and $T_{ll}$. It is inappropriate to claim that the limit \eqref{comlimCovApp} implies $T_{ab}\approx T_{kk}l_a l_b$, so Eq.\eqref{EMTintrnearhorizon} is not a general consequence of Eq.\eqref{EMTintrCommonLimit}.
The situation for $\theta_{l}=0$ is analogous, as confirmed by simply swapping $k^a$ and $l^a$. In the next section, we will show that the Einstein tensor of the Ellis drain hole provides a tractable example demonstrating that Eq.\eqref{EMTintrnearhorizon} does not generally hold.

%%%%%%%%%%%%%%%%%%%%%%%%%%%%
\section{Analytic example: Ellis drainhole spacetime}
\setcounter{equation}{0}
\renewcommand\theequation{3.\arabic{equation}}
The Ellis drainhole spacetime is a solution to the Einstein-phantom theory \cite{Ellis:1979bh}.
This spacetime offers a tractable analytical example that illustrates the details discussed in the previous section. We first introduce the metric and then compute the Einstein tensor\footnote{Appendix B gives details for the matter source supporting the Ellis drain hole} to demonstrate that Eq.\eqref{EMTintrnearhorizon} is not a physical result but arises from a coordinate singularity. The line element for the Ellis drainhole is
\begin{align}
	ds^2 = -d\eta^2 + d\rho^2 + r^2(\eta,\rho) d\Omega^2 \,,
\end{align}
where
\begin{align}
	r^2(\eta,\rho) = \alpha^2 \eta^2 + (1+\alpha^2) \rho^2  \,.  \label{rsqFunc}
\end{align}

%%%%%%%%%%%%% Identify trapping horizons  %%%%%%%%%%%%%%%%%%%%%%%%%%%%%%%%%%%%%%%%%%%%%%%%%%%%%%
The time-radial subspace is an explicit two-dimensional Minkowski spacetime. Thus, the following two vector fields 
\begin{align}
	k^{a}= \frac{1}{\sqrt{2} } \bigg(\frac{\partial u^a}{\partial \eta } + \frac{\partial u^a}{\partial \rho }\bigg) \,, \quad
	l^{a}= \frac{1}{\sqrt{2} } \bigg(\frac{\partial u^a}{\partial \eta } - \frac{\partial u^a}{\partial \rho }\bigg) \,,
\end{align}
describe null geodesics propagating along the radial direction. We have oriented these two null vector fields to point toward the future. Lowering their indices, we obtain
\begin{align}
	k_{a}  = \frac{1}{\sqrt{2} } (- \barnab_{a}\eta + \barnab_{a} \rho ) \,, \quad
	l_{a}  = \frac{1}{\sqrt{2} } (- \barnab_{a}\eta - \barnab_{a} \rho ) \,.
\end{align}
We further compute their expansions as
\begin{align}
	\theta_k 
	= \sqrt{2} \frac{ \alpha^2 \eta + (1+\alpha^2) \rho }{\alpha^2 \eta^2 + (1+\alpha^2) \rho^2}
	\,, \quad
	\theta_l = \sqrt{2} \frac{ \alpha^2 \eta - (1+\alpha^2) \rho }{\alpha^2 \eta^2 + (1+\alpha^2) \rho^2} \,,
\end{align}
Hence, the trapping horizons defined by $\theta_k=\theta_l=0$ satisfy the following equations
\begin{align}
	\rho = -\frac{\alpha^2}{1+\alpha^2} \, \eta
	\,, \quad
	\rho = \frac{\alpha^2}{1+\alpha^2} \, \eta \,,
\end{align}
both of which are timelike hypersurfaces.

%%% the Kodama vector field %%%
We next express $\barnab^a r$ in terms of $k_a$ and $l_a$:
\begin{align}
	\barnab_{a} r	=&~ \frac{\alpha^2 \, \eta +(1+\alpha^2)\, \rho }{\sqrt{2}\, r} l_{a} + \frac{-\alpha^2 \, \eta +(1+\alpha^2)\, \rho}{\sqrt{2}\, r} k_{a}
	\,.
\end{align}
Furthermore, the Kodama vector field is given by
\begin{align}
	K_{a} =&~ -\frac{(1+\alpha^2)\, \rho}{r}  \nabla_{a}\eta - \frac{\alpha^2 \, \eta}{r}  \nabla_{a}\rho   \nonumber \\
	=&~ \frac{\alpha^2 \, \eta +(1+\alpha^2)\, \rho }{\sqrt{2}\, r} l_{a} + \frac{\alpha^2 \, \eta -(1+\alpha^2)\, \rho}{\sqrt{2}\, r} k_{a}   \,.
\end{align}
Explicitly, $K^a = \barnab^a r$ on the horizon $\rho = \alpha^2\eta/(1+\alpha^2)$ and $K^a = -\barnab^a r$ on the horizon $\rho = -\alpha^2\eta/(1+\alpha^2)$. This confirms the relation $[K^a]_H = \pm [\barnab^a r]_H$.

%%% Components of the Einstein tensor  %%%
We further compute the components of the Einstein tensor as follows:
\begin{align}
	G_{\eta\eta} =& G_{\rho\rho} = -\frac{\alpha^2 (1+\alpha^2) (\rho^2+\eta^2) }{r^4} \,, %\nonumber\\ 
	\quad  G_{\eta \rho} = \frac{ 2\alpha^2 (1+\alpha^2) \rho \eta }{r^4} \,.
\end{align}
On the trapping horizons, these components reduce to
\begin{align}
	G_{\eta\eta}|_H =&~ G_{\rho\rho}|_H = - \frac{(1+\alpha^2)(1+2\alpha^2+2\alpha^4)}{\alpha^2(1+\alpha^2)^2} \frac{1}{\eta^2} \,, \label{Einetaeta}\\ 
	G_{\eta \rho}|_H =& \pm \frac{2(1+\alpha^2)^2}{(1+2 \alpha^2)^2} \frac{1}{\eta^2} \label{Einetarho}\,,
\end{align}
where the $+$ sign corresponds to $\rho = \alpha^2\eta/(1+\alpha^2)$ and the $-$ sign to $\rho = -\alpha^2\eta/(1+\alpha^2)$.
Clearly, Eq.\eqref{Einetaeta} contradicts Eq.\eqref{Einetarho}. Consequently, Eq.\eqref{EMTintrnearhorizon} is not satisfied in a coordinate system free of coordinate singularities.

%%%%%%%%%%%%%%%  Construct \texorpdfstring{$\{t,r\}$}{t,r} coordinates   %%%%%%%%%%%%%%%%%%%%%%%%%%%%%%%%%%%%%%%%%%%%%%% 
We next seek the coordinate transformation from $\{\eta, \rho\}$ to $\{t, r\}$. The key step is to identify the function $t$ that labels slices orthogonal to the constant-$r$ hypersurfaces. We recall that  
\begin{align}
	\barnab_{a} r =&~  \frac{\alpha^2 \, \eta}{r} \barnab_{a}\eta + \frac{(1+\alpha^2)\, \rho}{r} \barnab_{a}\rho  \,.
\end{align}
We compute $f = \gbar^{ab} \barnab_{a} r \barnab_{b} r$ as
\begin{align}
	f =&~ -\bigg( \frac{\alpha^2 \, \eta}{r} \bigg)^2 + \bigg( \frac{(1+\alpha^2)\, \rho}{r} \bigg)^2  \nonumber\\
	=&~ \frac{1}{r^2} \big( (1+\alpha^2) \rho - \alpha^2 \eta \big) \big( (1+\alpha^2) \rho + \alpha^2 \eta \big)
	\,.  \label{grrFunc}
\end{align}
Then the function labeled time slices orthogonal to constant $r$ can be constructed via $\gbar^{ab} \barnab_{a} t \barnab_{b} r =0 $. Concretly, the orthogonal condiction implies the following exact one-form
\begin{align}
	\barnab_{a} t \propto &~  \frac{ \barnab_{a}\eta }{\alpha^2 \, \eta }  + \frac{ \barnab_{a}\rho }{(1+\alpha^2)\, \rho} 
	\,. 
\end{align}
\begin{figure}[ht]%[htbp!][H]
	\begin{center}
		\includegraphics[width=0.6\textwidth]{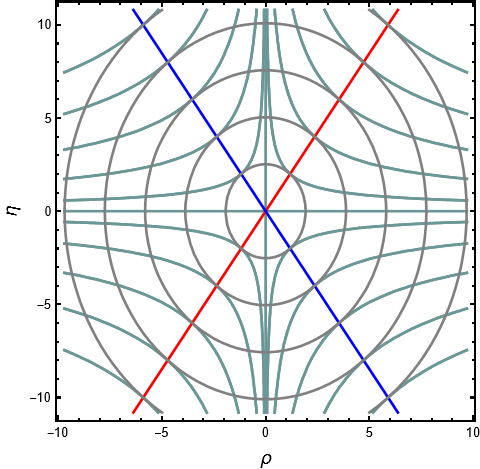}
	\end{center}
	\caption{\small We set $\alpha=1.2$.  Gray contours represent the surfaces of constant $r$, and muted teal contours represent those time slices $t$ that are orthogonal to the surfaces of constant $r$. The red line denotes the trapping horizon $\theta_k=0$, while the blue line denotes the trapping horizon $\theta_l=0$. On these horizons, the $t$-slices are tangent to the surfaces of constant $r$, as their normal vectors become null. This implies that $dt$ and $dr$ become collinear, giving rise to the coordinate singularity in the  $\{t, r\}$ coordinate system.	} 
\label{rhoperaVSu}  
\end{figure}
To visualize key features, we plot the constant-$t$ and constant-$r$ contours in the $\eta$-$\rho$ plane for $\alpha=1.2$ in Fig.\ref{rhoperaVSu}.
If one chooses a function $t$ that covers a trapping horizon, for instance, the upper-right segment of the red line in Fig.\ref{rhoperaVSu}, the $t$ slice generally intersects each constant-$r$ hypersurface twice.
Notably, on the horizons, the normal vectors $\barnab^a t$ and $\barnab^a r$ are tangent to one another, consistent with the illustration in Fig.\ref{Ktangentrconst}.
We therefore divide the entire $\M2$ into four regions to ensure the validity of $\{t,r\}$ as a coordinate system:
\begin{align}
	t_\text{R} =&~ \big(w\rho\big)^{ \frac{\alpha^2}{1+\alpha^2} } \eta 
	\,, \quad
	t_\text{L} = \big(-w\rho\big)^{ \frac{\alpha^2}{1+\alpha^2} } \eta  \,, \nonumber\\
	t_\text{U} =&~ \big(w\eta\big)^{ \frac{1+\alpha^2}{\alpha^2} } \rho
	\,, \quad
	t_\text{D} = \big(-w\eta\big)^{ \frac{1+\alpha^2}{\alpha^2} } \rho  \,.
\end{align}

Then, it is beneficial to compute $e^{-2h} G_{tt}$, $\pm e^{-h}G^r_{\;t}$, and $G^{rr}$ explicitly to verify Eq.\eqref{EMTintrCommonLimit}.
Without loss of generality, we consider the right region with $t = \big(w\rho\big)^{ \frac{\alpha^2}{1+\alpha^2} }  \eta $, where the subscript $\text{R}$ is omitted. The line element in the $\{t,r,\theta,\phi\}$ coordinate system then becomes
\begin{align}
	ds^2 = -\frac{r^2}{t^2f}\frac{1+\alpha^2}{\alpha^2(1+2\alpha^2)^2} (\alpha^2 +f)(1+\alpha^2-f) dt^2 + \frac{dr^2}{f} + r^2 d\Omega^2 \,, \label{ds2intr}
\end{align}
where $f$ as a function of $t$ and $r$ is determined by
\begin{align}
	\frac{1+\alpha^2-f}{\alpha^2(1+2\alpha^2)} \bigg(\frac{\alpha^2+f}{1+3\alpha^2+2\alpha^4}\bigg)^{\frac{\alpha^2}{1+\alpha^2}} = \big(wt\big)^2 \big(wr\big)^{-2\frac{1+2\alpha^2}{1+\alpha^2}} \,.  \label{foftr}
\end{align}
We define $e^{2h} = -g_{tt}g_{rr}$ using the metric components $g_{tt} = - e^{2h} f$ and $g_{rr} = 1/f$. Solving for $e^h$ yields 
\begin{align}
	\pm e^{h} 	= \frac{1}{t} \frac{r}{f} \frac{\sqrt{1+\alpha^2}}{\alpha(1+2\alpha^2)}	\sqrt{ (\alpha^2 +f)(1+\alpha^2-f)}	\,.
\end{align}
We then compute the components of $G^{ab}$ in the $\{t,r\}$ coordinate system, utilizing the relations $ G^{\eta\eta} = G_{\eta\eta} $, $ G^{xx} = G_{xx} $, and $ G^{\eta x} = -G_{\eta x} $:
\begin{align}
	G^{tt} =& - \frac{\alpha^2(1+2\alpha^2)^2}{1+\alpha^2}  \frac{2\alpha^2(1+\alpha^2)+f(1-f)}{(1+\alpha^2-f)(\alpha^2+f)} \frac{t^2}{r^4} \,,\nonumber\\
	G^{rr} =& - \frac{2  \alpha^2 (1 + \alpha^2)}{r^2} - \frac{f(1-f)}{r^2}  \,,\nonumber\\
	G^{rt} =& - 2  \alpha^2 (1 + 2\alpha^2)\frac{t}{r^3} \,,
\end{align}
which verify the following relations:
\begin{align}
	e^{-2h}G_{tt} =&~ e^{2h} f^2 G^{tt} \nonumber\\
	=& - \frac{2  \alpha^2 (1 + \alpha^2)+f(1-f)}{r^2} =	G^{rr}  \,, \nonumber\\
	%	\nonumber\\
	\pm e^{-h} G^{r}_{\;\,t} =& \mp e^{h} f G^{rt} \nonumber\\
	=&~ \frac{2}{r^2} \sqrt{ (\alpha^4 + \alpha^2 f)[(1+\alpha^2)^2-(1+\alpha^2)f]}	\,.  \label{EinforEllisintrRight}
\end{align}
Setting $f=0$ gives $e^{-2h} G_{tt} = G^{rr}$ and $[e^{-h}G^r_{\;t}]_H = \pm [G^{rr}]_H$, which are consistent with the general limit in Eq.\eqref{EMTintrCommonLimit}.
The validity of Eq.\eqref{EMTintrCommonLimit} does not imply the validity of Eq.\eqref{EMTintrnearhorizon}. Additionally, Eq.\eqref{foftr} yields
\begin{align}
	\frac{\partial f}{\partial r}  =  \frac{2 (1+\alpha^2-f)(\alpha^2+f) }{rf}  \,.
\end{align}
The expression for $\partial C/\partial r$ in terms of $r$ and $f$ is thus
\begin{align}
	\frac{\partial C}{\partial r}  =&  -\frac{2\alpha^2(1+\alpha^2)}{f} -1+f  \nonumber\\
	\simeq& -\frac{2\alpha^2(1+\alpha^2)r_H}{r-r_H} = \frac{8\pi\Xi\,r_H^3}{r-r_H} \,,
\end{align}
where we have used $8\pi\Xi = [G^{rr}]_H = -2\alpha^2 (1+\alpha^2)/r_H^2 < 0$ from Eq.\eqref{EinforEllisintrRight}.
This confirms Eq.\eqref{MMSpriintx}.
We perform the same calculation for $\partial h/\partial r$:
\begin{align}
	\frac{\partial h}{\partial r}  =& -\frac{2\alpha^2(1+\alpha^2)+f(1-f)}{rf^2} = \frac{1}{rf} \frac{\partial C}{\partial r}  \nonumber\\
	\simeq& -\frac{2\alpha^2(1+\alpha^2)r_H}{(r-r_H)^2} =\frac{8\pi\Xi\,r_H^3}{(r-r_H)^2} \,,
\end{align}
which confirms the result in Eq.\eqref{hpriintx}.

\section{Conclusion and discussion}
\setcounter{equation}{0}
\renewcommand\theequation{4.\arabic{equation}}

%summary
We clarify the physical picture underlying the coordinate singularity induced by $f\rightarrow 0$ in the orthogonal coordinate system $\{t,r\}$. This picture demonstrates that the trapping horizon gives rise to a coordinate singularity in the $\{t,r,\theta,\phi\}$ coordinate system for dynamically spherical spacetimes. Building on this picture, we rederive the fundamental results widely applied in a series of works concerning physical black holes and identify the misinterpreted features of the energy-momentum tensor (EMT). For any second-order symmetric tensor $T_{\mu\nu}$ that is regular on the horizon, the components $e^{-2h}T_{tt}$, $T^{rr}$, and $e^{-2h}T^{r}_{\;t}$ approach the same value up to a sign. Nevertheless, we utilize a null tetrad to clarify that these three limits only capture one component of $T_{ab}$ on the horizon, while neglecting the other two components of $T_{ab}$. It is therefore incorrect to assert that $T_{ab}$ exhibits a universal structure on horizons. This necessitates a re-evaluation of other conclusions derived in the series of works \cite{Baccetti:2018qrp, Baccetti:2019mab, Murk:2020wkm, Dahal:2021iez, Terno:2019kwm, Terno:2020tsq, Dahal:2021hbm,Terno:2022qot}.

The Ellis drain hole serves as an optimal example to verify that Eq.\eqref{EMTintrnearhorizon} is not universally valid. One advantage lies in the ease of constructing a concrete analytic expression for the coordinate transformation to the $\{t,r\}$ coordinate system. Consequently, the correctness of Eqs.\eqref{EMTintrCommonLimit}, \eqref{MMSpriintx}, and \eqref{hpriintx} is confirmed in a highly traceable manner. This explicitly indicates that the near-horizon structure of the EMT is not a general consequence of the regularity condition, but rather an artifact of the coordinate singularity. In future work, it may be beneficial to apply the covariant approach to other analytic models in Refs.\cite{Yang:2021diz, Liang:2025lay} or the numerical model in Ref. \cite{Shinkai:2002gv}.

\begin{figure}[htbp]%[htbp!][H]
	\centering
		\subfigure[\label{TSforming}]{\includegraphics[width=0.4\textwidth]{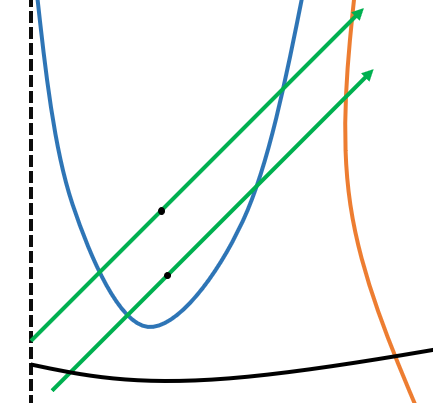}}
 \hspace{0.06\textwidth}
\subfigure[\label{ske_nullgeo}]{\includegraphics[width=0.4\textwidth]{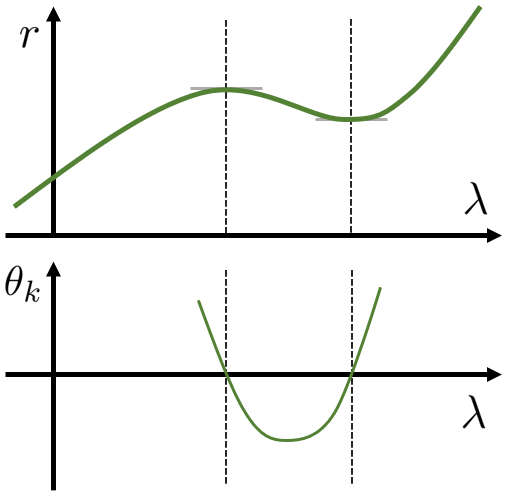}}
	\caption{\small (a) Black dots denote two trapped surfaces, and the vertical dashed line represents a regular center. The horizontal black curve corresponds to the hypersurface containing the initial data, while the orange vertical curve depicts a distant observer that never falls into the trapped region. Green arrows indicate two radial outgoing null geodesic congruences, and the blue curve denotes the trapping horizon. The visibility of the trapping horizon necessitates that the null geodesics attain a local minimum in areal radius before reaching the observer. When extending the null geodesics into the past, they may either be captured by the initial hypersurface or converge to the regular center. (b ) The variation of $r$ and $\theta_{k}$ with the affine parameter $\lambda$ are illustrated. We assume $\lambda$ increases along the future direction of the outgoing null geodesic congruence. Since $\theta_{k}$ has two roots, one corresponds to the local minimum of $r$ (on the outer part of the horizon), and the other to the local maximum (on the inner part of the horizon).
	} 
	\label{visibleTSforming}  
\end{figure}

Returning to the original motivation, it is worth noting that the visible trapped surface requirement constitutes another aspect of Theorem 9.2 in Hawking \& Ellis \cite{HawkingEllis}, as well as Theorems 12.2.2 and 12.2.3 in Ref. \cite{WaldGR}. Additionally, if we assume a visible trapped surface forms from an initial hypersurface without trapped surfaces, it is natural to expect that the areal radius of certain outgoing null geodesics first reaches a local maximum, then a local minimum, and ultimately increases again as these geodesics escape to large distances. This scenario is illustrated in Fig.\ref{visibleTSforming}.
In particular, as a consequence of visible trapped surfaces for distant observers, the areal radius along outgoing null geodesics must exhibit a local minimum, i.e., $[d^2r/d\lambda^2]_H>0$. This behavior corresponds to a violation of the null energy condition (NEC) since $\Xi<0$. Conversely, the NEC is satisfied when the areal radius along the null geodesics has a local maximum, which typically occurs within the infalling matter. This can also be inferred from the behavior of $\theta_{k}$, see Fig.\ref{ske_nullgeo}. Recall that $\theta_{k} \propto dr/d\lambda$, with $\theta_{k}$ positive outside the horizon and negative inside it. Along outgoing null geodesics, $\theta_{k}$ first decreases (i.e., $d \theta_{k}/d\lambda\propto d^2r/d\lambda^2<0$) and then increases (i.e., $d \theta_{k}/d\lambda\propto d^2r/d\lambda^2>0$). This implies that $\Xi>0$ and the NEC is satisfied when the outgoing null geodesic crosses the trapping horizon for the first time, whereas $\Xi<0$ and the NEC is violated upon the second crossing of the trapping horizon.

Furthermore, higher-order scenarios satisfying $\Xi \sim \mathcal{O}(f^s)$ with $s>1$ are likely important for horizon formation and evolution. As illustrated in Fig.\ref{visibleTSforming}, there exists a location on the horizon tangent to the outgoing null geodesic congruence. One might still anticipate that higher-order scenarios would yield a physical picture distinct from that presented in Fig.\ref{Ktangentrconst}. However, from another perspective, higher-order scenarios appear to be less problematic. Suppose $r\sim (\lambda-\lambda_c)^{s}$ asymptotically, where $\lambda_c$ corresponds to the position of the MTS. In this case, $\theta_{k} \sim s(\lambda-\lambda_c)^{s-1}$ and $\Xi \sim s(s-1)(\lambda-\lambda_c)^{s-2}$. A higher-order scenario implies $s>2$, meaning $\Xi$ also tends to zero as $f\rightarrow 0$, but the regularity of the horizon ensures that the second term in Eq.\eqref{EinTensor_rr} decays more rapidly than the first term. Nevertheless, we leave the investigation of higher-order scenarios to future work.

\setcounter{secnumdepth}{0} 

\section*{Acknowledgments}
J.Yang is supported by the Guangzhou Municipal Postdoctoral Research Project Funding, the National Natural Science Foundation of China (Grant No. 11873025 and Grant No.12133004), and the National SKA Program of China (Grant No. 2020SKA0120101). H. Huang further gratefully acknowledge support by the
National Natural Science Foundation of China (Grant No. 12565010). W-C.G. is supported by the National Natural Science Foundation of China under the Grant No. 12405064, Jiangxi Provincial Natural Science Foundation under the Grant No. 20242BCE50055, and the Initial Research Foundation of Jiangxi Normal University.

\section{Appendix A: Useful results}
\setcounter{equation}{0}
\renewcommand\theequation{A.\arabic{equation}}

\renewcommand\theequation{A.\arabic{equation}}

This appendix summarizes key results for the Levi-Civita connection and Riemann curvature tensor in an $n$-dimensional spacetime with the line-element:
\begin{align}
	ds^2 = \gbar_{ab} du^a du^b + r^2(u) \ghat_{ij}d\theta^i d\theta^j  \,,
\end{align}
as contextualized earlier. Then, the components of its inverse metric are given by
\be
g^{ab}=\gbar^{ab} \,,\quad g^{ij}=\fft{\ghat^{ij}}{r^2} \,.
\ee
Each component of the Levi-Civita connection is explicitly expressed as 
\begin{align}
	&\Gamma^a_{\;bc}
	=\bar{\Gamma}^a_{\;bc} \,,\;\;\nonumber\\
	&\Gamma^a_{\;ij} = -r \gbar^{ab}\fft{\partial r}{\partial u^b}\, \hat{g}_{ij} \,,\;\;\nonumber\\
	&
	\Gamma^i_{\;aj} = \fft{1}{r} \fft{\partial r}{\partial u^a} \delta^i_j\,,\;\;\nonumber\\
	&
	\Gamma^i_{\;jk}  =\hat{\Gamma}^i_{\;jk}  \,. \label{dredGam}
\end{align}
Notably, the components $\Gamma^a_{\,bc}$ and $\Gamma^i_{\,jk} $ correspond exactly to the independent Levi-Civita connections of $\M2$ and $\Mco2$ respectively.
We introduce the covariant differential operators $\bar{\nabla}_{a}$ for $\M2$ and $\hat{\nabla}_{i}$ for $\Mco2$. The areal radius $r(u)$ is treated as a scalar field on $\M2$, where $\barnab_a r$ denotes $\del r/\del u^a$ and $\barnab^a r$ is defined as $I^{ab}(\del r/\del u^b$).
The components of the Riemann tensor are derived as
\begin{align}
	&R^a_{\;bcd} = \bar{R}^a_{\;bcd} \,,\qquad  \nonumber\\
	&R^a_{\;ibj} = -R^a_{\;ijb} = -r (\barnab^a \barnab_b r)\,\ghat_{ij} \,,\;  \nonumber\\
	&R^i_{\;ajb} = -R^i_{\;abj} = -\fft{\barnab_a \barnab_b r}{r}\delta^i_j \,, \quad  \nonumber\\
	&
	R^i_{\;jkl} = \hat{R}^i_{\;jkl}- f(\delta^i_{\;k}\hat{g}_{jl}-\delta^i_{\;l}\hat{g}_{jk})\,. \label{dimreduc}
\end{align}
This result is consistent with those reported in Refs.~\cite{Maeda:2007uu, Yang:2023nnk}.
We further assume the transverse space has constant curvature, leading to
\begin{align}
  \hat{R}^i_{\;jkl} = \mathcal{K} (\delta^i_{\;k}\hat{g}_{jl}-\delta^i_{\;l}\hat{g}_{jk})\,,
\end{align}
 Contracting with $\delta^i_{\;i}=n-2$, the components of the Ricci tensor are obtained as
\begin{align}
	&R_{ab}=\bar{R}_{ab}-(n-2)\frac{\bar{\nabla}_a \bar{\nabla}_b r}{r} \;,\; \nonumber\\
	&	R_{ij}=\hat{g}_{ij}\big[(n-3) (\mathcal{K}-f)-r\bar{\nabla}^2 r
 \big] 
\;,	\label{dimreducRic}
\end{align}
where $\bar{\nabla}^2 r$ is $\bar{\nabla}^a \bar{\nabla}_a r$ for short. The Ricci scalar is given by
\begin{equation}
	\begin{split}
		R &= \bar{R} -2(n-2)\frac{\bar{\nabla}^2 r}{r}
		 -(n-2)(n-3)\frac{ \mathcal{K}-f }{r^2} \,.\label{RicciSca}
	\end{split}
\end{equation} 
This work focuses on 2-dimensional $\M2$. Since all 2-dimensional metrics are conformally flat (see \cite{WaldGR}), the Einstein tensor of a 2-dimensional metric vanishes identically, i.e., $\bar{R}_{ab}-\Rbar I_{ab}/2=0$. 
For simplicity, we define $\cK=\Rhat/((n-2)(n-3))$.
The Einstein tensor $G_{\mu\nu}=R_{\mu\nu}- Rg_{\mu\nu}/2$ thus takes the form:
\begin{align}
	G_{ab}=& -\frac{n-2}{r}\, \barnab_a \barnab_b\, r 
	+\frac{n-2}{r} \gbar_{ab}\,\bigg(\barnab^2r -\frac{n-3}{2r}(\mathcal{K}-f) \bigg) \nonumber \;,\\
	G_{ij}=& -\hat{g}_{ij}\bigg( \frac{r^2}{2} \Rbar-(n-3)\, r\barnab^2 r
	+\frac{(n-3)(n-4)}{2}(\mathcal{K}-f) \bigg) \;, %\label{dimreducEin} 
\end{align} 
For $n=4$, we compute the quantities $G_{ab}K^a K^b $, $G_{ab}K^a \barnab^br $ and $G_{ab}\barnab^ar  \barnab^br $. For comparison, the results are presented below:
%%%%%%%%%%%%%%%%%%%%%%%%%%%%%%%%%%%%%%%%
\begin{align}
	&G_{ab} K^aK^b = -\frac{2}{r}\,  \big(K^aK^b \barnab_a \barnab_b\, r \big)
	-f\,\bigg( \frac{2}{r}\barnab^2r -\frac{1-f}{r^2} \bigg) \nonumber \;,\\
	&G_{ab}(\barnab^a r) (\barnab^b r)= -\frac{2}{r}\, \big(\barnab^a r\big)\big(\barnab^b r\big) \big(\barnab_a \barnab_b\, r \big)
	+ f\,\bigg( \frac{2}{r}\barnab^2r -\frac{1-f}{r^2} \bigg) \nonumber \;,\\
	&G_{ab}K^a (\barnab^b r) = -\frac{2}{r}\, K^a\big(\barnab^b r\big)\big( \barnab_a \barnab_b\, r \big)   \;.
\end{align}
%%%%%%%%%%%%%%%%%%%%%%%%%%%%%%%%%%%%%%%%
 These expressions show that the terms proportional to $f$ vanish on the horizon. 
%%%%%%%%%%%%%%%%%%%%%%%%%%%%%%5 
\setcounter{secnumdepth}{0} 
\section{Appendix B: phantom configuration}
\setcounter{equation}{0}
\renewcommand\theequation{B.\arabic{equation}}

It is convenient to list the relevant properties of the matter source supporting the spacetime geometry of the Ellis drainhole. The matter field is a phantom scalar field without a potential term in the Lagrangian. Specifically, the full action reads
\begin{align}
  S = \int  \bigg( \frac{R}{16\pi} + \frac{1}{2}\nabla_{\mu}\phi \nabla^{\mu}\phi\bigg) \sqrt{-g}\, d^4x\,,
\end{align}
where the convention adopted here differs slightly from those in Refs.\cite{Ellis:1979bh, Yang:2021diz}. Hence, the equations of motion (EOM) are
\begin{align}
	&G_{\mu\nu} = 8\pi T_{\mu\nu} 
	= 8\pi\bigg( -\nabla_{\mu}\phi \nabla_{\nu}\phi +\frac{1}{2}g_{\mu\nu} \,\nabla_{\lambda}\phi \nabla^{\lambda}\phi \bigg) \,,\\
   &\nabla_{\lambda}\nabla^{\lambda}\phi = 0 \,.\label{Ein_phantomfreeEOMs}
\end{align}
The phantom field configuration for the Ellis drainhole is
\begin{align}
	\phi = \frac{1}{\sqrt{4\pi}} \arcsin\bigg(\sqrt{\frac{\alpha^2(1+\alpha^2)}{1+2\alpha^2}} \, \frac{\eta\pm\rho}{r}\bigg) \,. \label{phantom4EdrainH}
\end{align}
To confirm that this configuration satisfies the EOMs, we calculate its derivatives and find
\begin{align}
  \frac{\partial\phi}{\partial \eta} =  \sqrt{\frac{ \alpha^2(1+\alpha^2)}{4\pi}}\, \frac{\rho}{r^2} \,,\quad
  \frac{\partial\phi}{\partial \rho} = -\sqrt{\frac{ \alpha^2(1+\alpha^2)}{4\pi}}\, \frac{\eta}{r^2}  \,.
\end{align}
Thus, the phantom field satisfies
\begin{align}
 \nabla_{\lambda}\nabla^{\lambda}\phi = -\frac{\partial}{\partial\eta}\bigg(r^2\frac{\partial\phi}{\partial\eta}\bigg)	
 +\frac{\partial}{\partial\rho}\bigg(r^2\frac{\partial\phi}{\partial\rho}\bigg) =0 \,,
\end{align}
thereby satisfying the free Klein-Gordon equation.
Furthermore, the derivatives of the phantom field give
\begin{align}
	\nabla_{\lambda}\phi\nabla^{\lambda}\phi = \frac{ \alpha^2(1+\alpha^2)}{4\pi}\, \frac{\eta^2-\rho^2}{r^4}  \,.
\end{align}
We then compute all components of the EMT 
\begin{align}
	T_{\eta\rho}
	=&~ -\frac{ \alpha^2(1+\alpha^2)}{4\pi}  \frac{\eta\rho}{r^4}
	\,,\nonumber\\
	T_{\eta\eta} = T_{\rho\rho}
	=&~ -\frac{ \alpha^2(1+\alpha^2)}{8\pi}  \frac{\eta^2+\rho^2}{r^4}
	\,,\nonumber\\
	T_{ij} =&~ \ghat_{ij}\frac{ \alpha^2(1+\alpha^2)}{8\pi}\, \frac{\eta^2-\rho^2}{r^2} 	\,.
\end{align}
The components $T_{\eta\rho}$, $T_{\eta\eta}$ and $ T_{\rho\rho}$ match the corresponding components of the Einstein tensorr $G_{\eta\rho}$, $G_{\eta\eta}$ and $ G_{\rho\rho}$ provided in Section 3.
To complete our comparison, we present the $ij$ components of the Einstein tensor as follows:
\begin{align}
		G_{ij}=&~ r\hat{g}_{ij}\bigg(-\frac{\partial^2 r}{\partial\eta^2}+\frac{\partial^2 r}{\partial\rho^2}\bigg) \nonumber\\
		=&~ \alpha^2(1+\alpha^2) \frac{ \eta^2
			-\rho^2}{r^2}
		\,. %\label{dimreducEin} 
\end{align}
Therefore, Eq. \eqref{phantom4EdrainH} and the line element of the Ellis drain hole satisfy these EOMs \eqref{Ein_phantomfreeEOMs}.

%\newpage		
\bibliographystyle{JHEP}
\bibliography{lessRefs} 
\end{CJK}
\end{document}